\documentclass[aps,pra,showpacs,twocolumn,bibnotes]{revtex4}
\usepackage{bm}
\usepackage[T1]{fontenc}
\usepackage{mathrsfs}
\usepackage{natbib}
\usepackage{amsmath,amssymb}
\usepackage{latexsym}
\usepackage[dvips]{graphicx}
\begin{document}

\title{Molecular orbitals and strong-field approximation}
\author{Thomas Kim Kjeldsen}
\author{Lars Bojer Madsen}
\affiliation{Department of Physics and
Astronomy, University of Aarhus, 8000 {\AA}rhus C, Denmark}

\begin{abstract}
V.I. Usachenko and S.-I. Chu [Phys. Rev. A {\bf 71}, 063410 (2005)]
discuss the molecular strong-field approximation in the velocity
gauge formulation and indicate that some of our earlier velocity
gauge calculations are inaccurate. Here we comment on the results of
Usachenko and Chu. First, we show that the molecular orbitals used
by Usachenko and Chu do not have the correct symmetry, and second,
that it is an oversimplification to describe the molecular orbitals
in terms of just a single linear combination of two atomic orbitals.
Finally, some values for the generalized Bessel function are given
for comparison.
\end{abstract}

\maketitle
\section{Introduction}
\label{sec:Intro} The molecular strong-field approximation
(MO-SFA) was formulated in the velocity gauge in
Ref.~\cite{muthbohm00}, and more recently in the length
gauge~\cite{kjeldsen04a}. The latter work shows that the velocity
gauge MO-SFA with the initial state of the highest occupied
molecular orbital (HOMO) obtained in a self-consistent
Hartree-Fock (HF) calculations does not account for the observed
orientation-dependent ionization of N$_2$~\cite{corkum03}. Only
the corresponding length gauge MO-SFA~\cite{kjeldsen04a} and the
molecular tunneling theory~\cite{lin03} give the correct behavior.

The orientation dependence of the strong-field ionization of N$_2$
was revisited in Ref.~\cite{usachenko05}. There a simple model is
made for the $3\sigma_g$ HOMO in terms of a {\it single} linear
combination of atomic orbitals (LCAO) using two atomic $p$ orbitals.
Then the MO-SFA is applied in the velocity gauge to calculate the
ionization rate, and it is concluded that the velocity MO-SFA, with
such an approximation for the HOMO, is capable of predicting the
observed minimum~\cite{corkum03} in the rate for perpendicular
orientation of the internuclear axis with respect to the linear
polarization of the laser. Based on these findings it is
stated~\cite{usachenko05} that our earlier velocity gauge
calculations~\cite{kjeldsen04a} are inaccurate because of
inaccuracies in our (i) molecular orbital and (ii) generalized
Bessel function. Here we refute both these claims by giving details
of our orbitals and generalized Bessel functions. Additionally, we
show that apart from having the wrong symmetry, the model used for
the $3\sigma_g$ HOMO of N$_2$ in Ref.~\cite{usachenko05} is too
simple and we therefore believe that the agreement
in~\cite{usachenko05} with experiment is accidental. To account for
the molecular structure more LCAOs have to be included as, e.g., in
HF calculations. With such a self-consistent wave function, the
velocity gauge MO-SFA, indeed gives the wrong prediction for the
orientation-dependent ionization~\cite{kjeldsen04a}. Finally, we
present accurate values from~\cite{kjeldsen04a} of the generalized
Bessel function since also in this case some discrepancies exist
with Ref.~\cite{usachenko05} -- a discrepancy which can be
attributed to the difference in wavelength (we use 800~nm
light~\cite{kjeldsen04a}, in \cite{usachenko05} 795~nm light is
used~\cite{usachenkonote}).

\section{Molecular orbitals}
\label{sec:MO}

The idea of the LCAO approach to molecular orbital theory is to
identify the molecular orbitals in a basis of atomic orbitals. For
a homonuclear diatomic molecule, we write a molecular wave
function $\Psi(\bm r)$ as a superposition of atomic orbitals which
are centered on each of the atomic cores. The nuclei are located
at the positions $\pm \bm R/2$ with respect to the center of the
molecule. The molecular orbitals are the eigenfunctions of the
Fock operator, i.e., of the kinetic energy operator and the full
HF potential. These eigenfunctions can be obtained by a
diagonalization of the matrix representation of the Fock operator
in the atomic centered basis, $\phi_{nlm}(\bm r \mp \bm R/2)$. A
great simplification of the problem is achieved if the basis
functions are symmetrized according to the molecular point group
($D_{\infty h}$).  Consider, e.g., the atomic $1s$
[$\phi_{100}(\bm r \mp \bm R/2)$] orbitals. A change of basis
leads to a new set of basis functions
\begin{eqnarray}
  \psi_{\sigma_{g} 1s}(\bm r) &=& \mathcal{N}_{\sigma_{g} 1s}
  [\phi_{100}(\bm r - \frac{\bm R}{2}) + \phi_{100}(\bm r + \frac{\bm R}{2})] \\
  \psi_{\sigma_{u} 1s}(\bm r) &=& \mathcal{N}_{\sigma_{u} 1s}
  [\phi_{100}(\bm r - \frac{\bm R}{2}) - \phi_{100}(\bm r + \frac{\bm
  R}{2})].
\end{eqnarray}
These new basis functions belong to definite symmetry
classifications in  $D_{\infty h}$, namely $\Sigma_g^+$  ('+'
combination) and $\Sigma_u^+$ ('-' combination). The 2s, 3s, etc.
orbitals are symmetrized similarly. If we extend the basis to
contain atomic $2p$ orbitals, the properly symmetrized basis
functions in the $\Sigma$ blocks are
\begin{eqnarray}
  \psi_{\sigma_{g} 2p}(\bm r)
  &=&\mathcal{N}_{\sigma_{g} 2p}[\phi_{210}(\bm r -
  \frac{\bm R}{2}) - \phi_{210}(\bm r + \frac{\bm R}{2})]\\
  \psi_{\sigma_{u} 2p}(\bm r)
  &=&\mathcal{N}_{\sigma_{u} 2p}[\phi_{210}(\bm r -
  \frac{\bm R}{2}) + \phi_{210}(\bm r + \frac{\bm R}{2})],
\end{eqnarray}
for the $2 p_0$ ($2p_z$) orbitals, while the `+' and `--'
combination of the $2p_{\pm 1}$ are of $\pi_u$ and $\pi_g$
symmetry, respectively, and hence not of interest in our present
discussion focussing on the $3\sigma_g$ HOMO of N$_2$.

It is important to note that for a quantitative HF calculation, the
basis is further extended until convergence is obtained. We may
easily check the correct parity ($g$ or $u$) of the basis function
by letting $\bm r \rightarrow -\bm r$ and using the fact that the
parity of $\phi_{nlm}$ is $(-1)^l$. We emphasize this point since in
Ref.~\cite{usachenko05}, the `+' combination of the $2p_z$ orbitals
is erroneously referred to as the gerade state. Although not of
relevance for our fully numerical HF calculations, we note that in
the hydrogenic basis used in Ref.~\cite{usachenko05} the Fourier
transform of the antisymmetric combination of atomic $2p_z$ orbitals
is
\begin{eqnarray}
    \tilde{\psi}_{\sigma_g 2p}(\bm p) =\mathcal{N}_{\sigma_g 2p}\sin\left(\bm{p \cdot
  R}/2\right) \frac{p}{(1+4p^2)^3}Y_{10}(\hat{\bm p}),
  \label{eqn:ft3sigmag}
\end{eqnarray}
\begin{figure}
\includegraphics[width=\columnwidth]{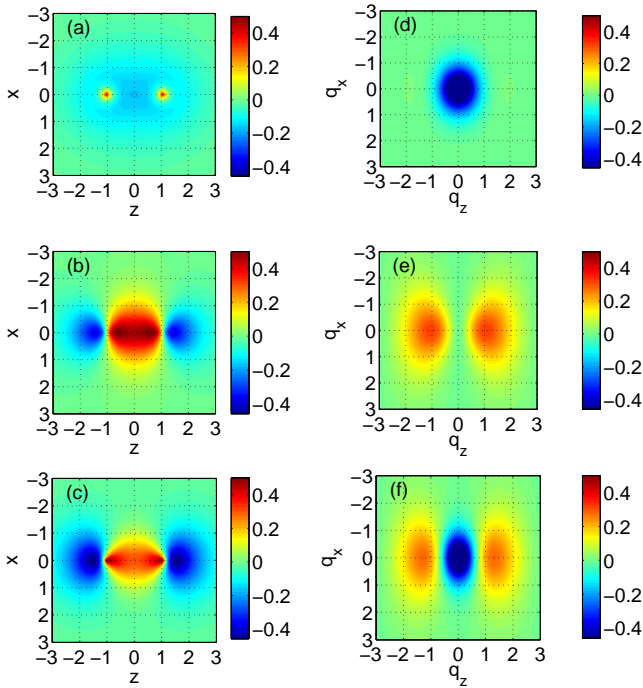}
\caption{(Color online). Planar cuts through the $3\sigma_g$ HOMO of
N$_2$ obtained in a HF calculation with a $(12s7p)/[6s4p]$ basis
(see text). Left (right) column: coordinate (momentum) space wave
function. Upper row: contribution from $s$ type basis functions,
middle row: contribution from $p$ type basis functions, lower row:
total HF result in the $(12s7p)/[6s4p]$ basis.}
  \label{fig:n2homo}
\end{figure}
with $\mathcal{N}_{\sigma_g 2p}$ the normalization constant. We may
verify the gerade parity symmetry by letting $\bm p \rightarrow -\bm
p$ and using $Y_{lm}(-\hat{\bm p}) = (-1)^l Y_{lm}(\hat{\bm p})$.
Equation~\eqref{eqn:ft3sigmag} is the correctly symmetrized version
of Eq.~(20) of Ref.~\cite{usachenko05}. The difference lies in the
sine-factor of Eq.~\eqref{eqn:ft3sigmag} compared with a
cosine-factor in Ref.~\cite{usachenko05}. A calculation with the
$p$-type basis functions alone would, because of this sine-factor,
predict suppressed ionization of N$_2$ compared with the companion
Ar atom -- in contrast to experimental and theoretical
findings~\cite{talebpour96,Guo98,muthbohm00}. Thus having pointed
out the importance of working with properly symmetrized basis
function, we now confront the {\it ad hoc} single LCAO
approch~\cite{usachenko05} with our HF calculations. Clearly, but
contrary to the approach in Ref.~\cite{usachenko05}, one cannot
expect the basis functions themselves to be eigenfunctions, instead
the eigenfunctions are linear combinations of basis functions.

\section{Hartree-Fock calculations}
\label{sec:HF}

In Ref.~\cite{kjeldsen04a} we used the GAMESS~\cite{gamess} program
for our quantitative HF calculations of N$_2$ with a
$(11s6p)/[5s3p]$ Cartesian Gaussian basis~\cite{dunning71}. The way
we specify our basis is standard in the quantum chemistry
literature~\cite{atkins}, but is recalled here for completeness. The
wave function is expanded in 11 $s$ and 6 $p$ orbitals centered on
each atom. These basis orbitals all have different exponentials in
the radial parts.  The orbitals of our HF calculation are obtained
by a variation of the expansion coefficients until the energy is
minimized. This procedure is equivalent to an iterative
diagonalization of matrix representations of the Fock operator. With
little loss of accuracy but significant gain in computational speed,
the optimization procedure can be simplified with a contracted basis
set.  This means that all the expansion coefficients are not varied
independently, contrary a fixed relationship is kept between some of
the coefficients. Such a simplification is used in most quantum
chemistry software.  Five coefficients corresponding to $s$ orbitals
and three coefficients corresponding to the $p$ orbitals are
optimized independently.  In order to describe the asymptotic
behavior we augment the basis by adding an extra diffuse $s$ and $p$
orbital. In total, the wave function is effectively expanded in a
basis of six $s$ and four $p$ orbitals on each atom denoted by
$(12s7p)/[6s4p]$ basis.

In Fig.~\ref{fig:n2homo} we show results of our calculations for the
$3\sigma_g$ HOMO of N$_2$ [Fig.~\ref{fig:n2homo}~(c)] and in its
decomposition of $s$ [Fig.~\ref{fig:n2homo}~(a)] and $p$
[Fig.~\ref{fig:n2homo}~(b)] type basis functions. We see that the
orbital is composed of both types of basis functions. At first sight
the $p$ contribution alone seems to resemble the orbital quite well.
However, in the proximity of the nuclei, it is the $s$ contribution
that dominates, since the $p$ functions have nodes on the nuclei. It
is the $s$ type contribution that dominates the momentum space wave
function at low momenta. This fact becomes clear in
Figs.~\ref{fig:n2homo}~(d)-(f), where we show the Fourier transforms
of the functions in Figs.~\ref{fig:n2homo}~(a)-(c). It is clear that
both the $s$ and $p$ type of basis functions contribute
significantly to the momentum wave function. Therefore, as discussed
in Sec.~\ref{sec:MO}, it is insufficient to consider each type
exclusively: The $s$ and $p_z$ basis orbitals are not separately
eigenfunctions of the Fock operator, contrary, the mixture turns out
to be essential. At this point the authors of
Ref.~\cite{usachenko05} introduce an oversimplification by
considering the HOMO to be constructed only from a single linear
combination of two $p$-orbitals.

\section{Implications to the molecular strong-field approximation}
\label{sec:SFA} In the MO-SFA one calculates the transition rate
from the HOMO to a continuum Volkov state through absorption of $n$
photons. For a linearly polarized, monochromatic field of frequency
$\omega$ and field strength $F_0$, the velocity gauge result for the
angular differential ionization rate in atomic units
is~\cite{muthbohm00}
\begin{eqnarray}
  \nonumber
  \frac{dW}{d\boldsymbol{\hat p}}
  &=& 2\pi N_e p_n \left| {\tilde \Psi}( {\bm p}_n) \right|^2(U_p-n\,\omega)^2\\
  &\times& J_n^2\left( \frac{|\bm F_0 \cdot \boldsymbol{p}_n|}{\omega^2},-\frac{U_p}{2\omega}\right),
  \label{eqn:diffrate}
\end{eqnarray}
with $N_e$ the number of electrons occupying the HOMO, $p_n =
\sqrt{2 (n\omega-U_p-I_p)}$ the momentum, $U_p = F_0^2/(4\omega^2)$
the ponderomotive energy, $I_p$ the ionization potential of the HOMO
($0.5725$ in the case of N$_2$~\cite{huber}), $\tilde{\Psi}( \bm p)$
the momentum space wave function and $J_n(u,v)$ a generalized Bessel
function~\cite{reiss80}.

According to the discussion in Sec.~\ref{sec:HF}, it is important to
apply the correct momentum space wave function in the evaluation of
Eq.~\eqref{eqn:diffrate}. To illustrate this fact, we consider the
total ionization rate integrated over all angles of the outgoing
electron and summed over all numbers of accessible photon
absorptions. We present this total ionization rate for different
angles $\theta$ between the internuclear axis and the polarization
axis for the HOMO $3\sigma_g$ orbital of N$_2$. We consider three
descriptions of the molecular orbital: (i) using $s$-type basis
functions only, (ii) using $p$-type basis functions only and (iii)
using the self-consistent HF solution. Clearly the approaches (i)
and (ii) only give a very poor description of the HOMO and therefore
these results deviate from the result obtained by using the true
HOMO.
\begin{figure}
  \begin{center}
  \includegraphics[width=0.8\columnwidth]{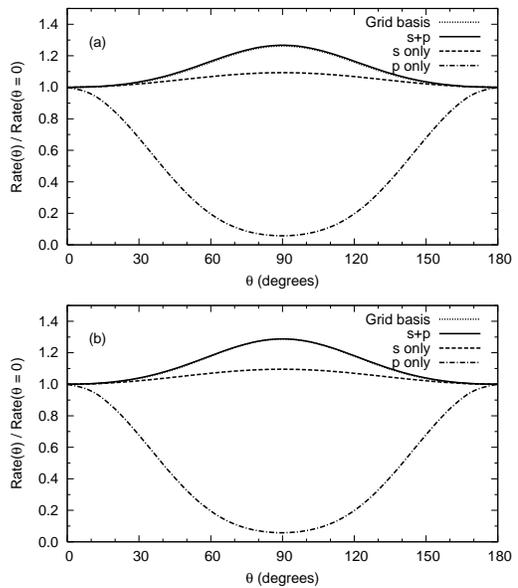}
 \end{center}
  \caption{Ionization rate at $2\times
  10^{14}\, \mbox{Wcm}^2$ of the HOMO of N$_2$ ($3\sigma_g$) as a function of the angle
  $\theta$  between the internuclear axis and the polarization axis at (a) 800~nm~\cite{kjeldsen04a} and
  (b) 795~nm~\cite{usachenkonote}. The
  short-dashed and solid curve are the calculations with the HF wave
  function derived from a grid calculation and with the $(12s7p)/[6s4p]$
  basis set in Fig.~\ref{fig:n2homo}~(f), respectively.
  The long-dashed curve with a maximum and the chained curve are
  obtained with
  the $s$ [Fig.~\ref{fig:n2homo}(d)] and $p$ [Fig.~\ref{fig:n2homo}(e)] contributions only, respectively.
  All rates are normalized to
  parallel $(\theta = 0)$ geometry.
  }
  \label{fig:ratetheta}
\end{figure}
In Fig.~\ref{fig:ratetheta} we present the results for the three
cases at (a) 800~nm~\cite{kjeldsen04a} and (b)
795~nm~\cite{usachenkonote}. Additionally, we show in
Fig.~\ref{fig:ratetheta} the results obtained with an initial HOMO
derived from a grid-based HF calculation~\cite{kobus96}. The
excellent agreement with the $(12s7p)/[6s4p]$ basis-state
calculation proves the convergence of the latter approach. We note
that panels (a) and (b) are very similar and hence the differences
between the results in~\cite{kjeldsen04a} and~\cite{usachenko05} are
not due to the slight difference in wavelength.

\begin{figure}
  \begin{center}
    \includegraphics[width=0.8\columnwidth]{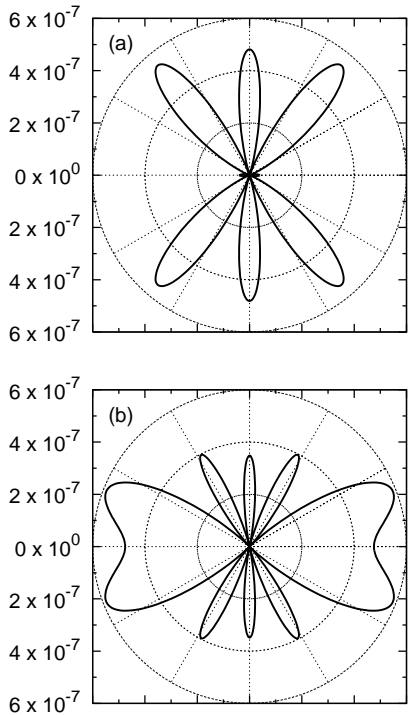}
  \end{center}
  \caption{Polar plot of the generalized Bessel function of order $n=18$
  squared for $2 \times 10^{14}$
  W/cm$^2$, and (a) 800~nm~\cite{kjeldsen04a}, (b)
  795~nm~\cite{usachenkonote}.  The polar
  angle $\theta_p$ is measured from the horizontal line.}
  \label{fig:bessel}
\end{figure}

From Fig.~\ref{fig:n2homo}~(d) we see that the momentum wave
function in case (i) is nearly spherically symmetric, and hence, in
Fig.~\ref{fig:ratetheta} the ionization rate in this case is nearly
independent of the molecular orientation. In case (ii) we use the
momentum function from Fig.~\ref{fig:n2homo}~(e). Although
exceptions occur, as evident from Fig.~\ref{fig:bessel}(a) below,
the generalized Bessel functions are typically maximized when $\bm
p$ is nearly parallel to the polarization direction
[Fig.~\ref{fig:bessel}(b)]. In a qualitative analysis, one may
therefore expect that the maximum rate is obtained, when the
polarization axis coincides with a direction where
$|\tilde{\Psi}(\bm p)|^2$ is large. Correspondingly, in case (ii) we
see from Fig.~\ref{fig:ratetheta} that the rate is maximized when
$\theta = 0$, i.e. when the polarization is horizontal in
Fig.~\ref{fig:n2homo}~(e) and overlaps with a large value of the
momentum wave function. Finally, we turn to the full wave function
in case (iii) which is the coherent sum of $s$ and $p$
contributions, Fig.~\ref{fig:n2homo}~(f). It is the low momenta that
contribute mostly to the total rate. At these small momenta, we see
from Fig.~\ref{fig:n2homo}~(f) that the absolute square of the
momentum wave function is maximized along the direction
perpendicular to the molecular axis and therefore the rate is
maximized with the molecule aligned perpendicularly to the laser
polarization, $\theta = \pi/2$. This result is also shown in
Fig.~\ref{fig:ratetheta} verifying the result of
Ref.~\cite{kjeldsen04a}. Note that the rate using the total wave
function is not the sum of rates of the $s$ and $p$ contributions
since their contributions to the wave function should be added
coherently and with the correct self-consistent amplitudes before
entering $\tilde{\Psi}( \bm p)$ in Eq.~\eqref{eqn:diffrate}. We note
that the admixture of atomic s and p  orbitals for the description
of the HOMO of N$_2$ was also recently pointed out in related work
on high-harmonic generation~\cite{Zimmermann05}.

In Refs.~\cite{kjeldsen04a,usachenko05}, the values of the
generalized Bessel function in Eq.~\eqref{eqn:diffrate} do also not
agree. Figure~\ref{fig:bessel}(a) shows our previous result for the
square of the generalized Bessel function corresponding to 18 photon
absorption at 800~nm and at an intensity of $2 \times 10^{14}$
W/cm$^2$. Panel~(b) shows the corresponding  result at
795~nm~\cite{usachenkonote}. Hence, Panel~(b) corresponds to
Fig.~2(a) of Ref.~\cite{usachenko05} (since $J_{-n}(u,v) =
(-1)^nJ_n(u,-v)$~\cite{reiss80}) and by comparison we see that the
results agree. We may therefore attribute the difference between the
generalized Bessel functions in \cite{kjeldsen04a} and
\cite{usachenko05} to stem from the difference in the wavelengths
used. It is clear from Fig.~\ref{fig:bessel} that the values of the
generalized Bessel function are very sensitive to the exact value of
the arguments.
We reproduce the values for the generalized Bessel function given in
Ref.~\cite{reiss03} with the input parameters used therein, and we
therefore believe that our previous and present calculations of the
generalized Bessel function are correct. Due to the excellent
agreement at 795~nm with the Bessel functions of
Ref.~\cite{usachenko05} we likewise believe that the latter are
evaluated correctly.

\section{Conclusion}
\label{sec:Con} In conclusion, the velocity gauge MO-SFA when based
on a transition to the continuum from an initial HOMO, determined by
a HF calculation using standard quantum chemistry software, gives
the wrong prediction for the orientation-dependent rate. In this
approximation only the length gauge MO-SFA (and the molecular
tunneling theory)~\cite{kjeldsen04a} give the observed
minimum~\cite{corkum03} in the rate for the perpendicular geometry.
This conclusion contrasts the findings in a recent
work~\cite{usachenko05}. In that work, however, only a single LCAO
of two $p$-orbitals (of wrong symmetry) was considered for the
description of the HOMO and no self-consistent wave function was
applied. From Fig.~\ref{fig:ratetheta} we clearly see, that an
application of a too simple wave function, e.g., one with only $p$
basis functions, leads to a prediction accidentally in qualitative
agreement with experiment~\cite{corkum03}. If one nevertheless would
be encouraged by this agreement to believe that the single LCAO of
$p$ orbitals describes the physics well, one should note that such a
wave function due to the presence of the sine-factor
in~\eqref{eqn:ft3sigmag} would predict suppressed ionization in
N$_2$ compared with Ar which is in contrast with experiments and
theory~\cite{talebpour96,Guo98,muthbohm00}. In closing we note that
also in the atomic case, the length gauge version of the SFA was
recently shown to be superior to the velocity gauge version when
compared with results obtained by integrating the time-dependent
Schr{\"o}dinger equation~\cite{Bauer05}.

\begin{acknowledgments}
  LBM is supported by the Danish Natural Science Research Council (Grant
No. 21-03-0163).
\end{acknowledgments}


\begin{thebibliography}{16}
\bibitem[{\citenamefont{first author of Ref.~\cite{usachenko05} has informed us
  that he used 795~nm light.}()}]{usachenkonote}
\bibinfo{author}{\bibfnamefont{The }~\bibnamefont{first author of
  Ref.~\cite{usachenko05} has informed us that he used 795~nm light}}.
\end{thebibliography}
\end{document}